\documentclass{article}
\usepackage{spconf,amsmath,graphicx}

\usepackage{hyperref}
\usepackage[ruled,linesnumbered]{algorithm2e}
\usepackage{amsfonts}
\usepackage{multirow}
\usepackage{graphicx}
\graphicspath{ {./figures/} }
\usepackage{caption}
\usepackage{subcaption}
\usepackage[bottom]{footmisc}
\usepackage{yonatan}
\usepackage{cite}

\title{MapiFi: Using Wi-Fi Signals to Map Home Devices}
\name{Yonatan Vaizman, Hongcheng Wang}
\address{Comcast --- Applied AI \& Discovery}
\begin{document}
%
\maketitle
\begin{abstract}
Imagine a map of your home with all of your connected devices (computers, TVs, voice control devices, printers, security cameras,~\etc), in their location. You could then easily group devices into user-profiles, monitor Wi-Fi quality and activity in different areas of your home, and even locate a lost tablet in your home. MapiFi is a method to generate that map of the devices in a home. The first part of MapiFi involves the user (either a technician or the resident) walking around the home with a mobile device that listens to Wi-Fi radio channels. The mobile device detects Wi-Fi packets that come from all of the home's devices that connect to the gateway and measures their signal strengths (ignoring the content of the packets). The second part is an algorithm that uses all the signal-strength measurements to estimate the locations of all the devices in the home. Then, MapiFi visualizes the home's space as a coordinate system with devices marked as points in this space. A patent has been filed based on this technology~\cite{mapifipatent}. This paper was published in~\cite{mapifiscte}.~\footnote{See published paper at \url{https://wagtail-prod-storage.s3.amazonaws.com/documents/SCTE_Technical_Journal_V1N3.pdf}.}
\end{abstract}
\section{Introduction}
\label{sec:intro}
Today's Wi-Fi access points (home-internet gateways) often come with applications to manage the home network and the devices connected to it (phones, computers, voice control devices, cameras,~\etc). Such applications may include features like creating user-profiles (assigning devices to people in the home), pausing internet access to certain devices (\eg~for parental control over a child's device), and Wi-Fi usage tracking (how much download/upload does each device take). These apps may also track Wi-Fi quality (\eg~channel interference, weak signals) and facilitate troubleshooting Wi-Fi problems for specific devices or for the overall home network.

\begin{figure}
     \centering
     \begin{subfigure}[t]{0.185\linewidth}
         \centering
         \includegraphics[width=\textwidth]{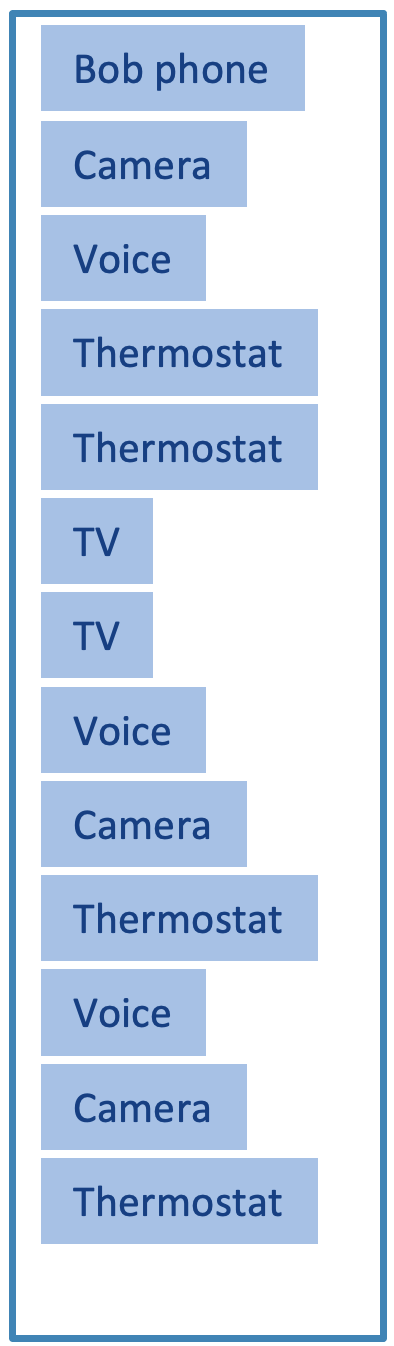}
         \caption{List}
         \label{fig:devlist}
     \end{subfigure}
    \begin{subfigure}[t]{0.80\linewidth}
         \centering
         \includegraphics[width=\textwidth]{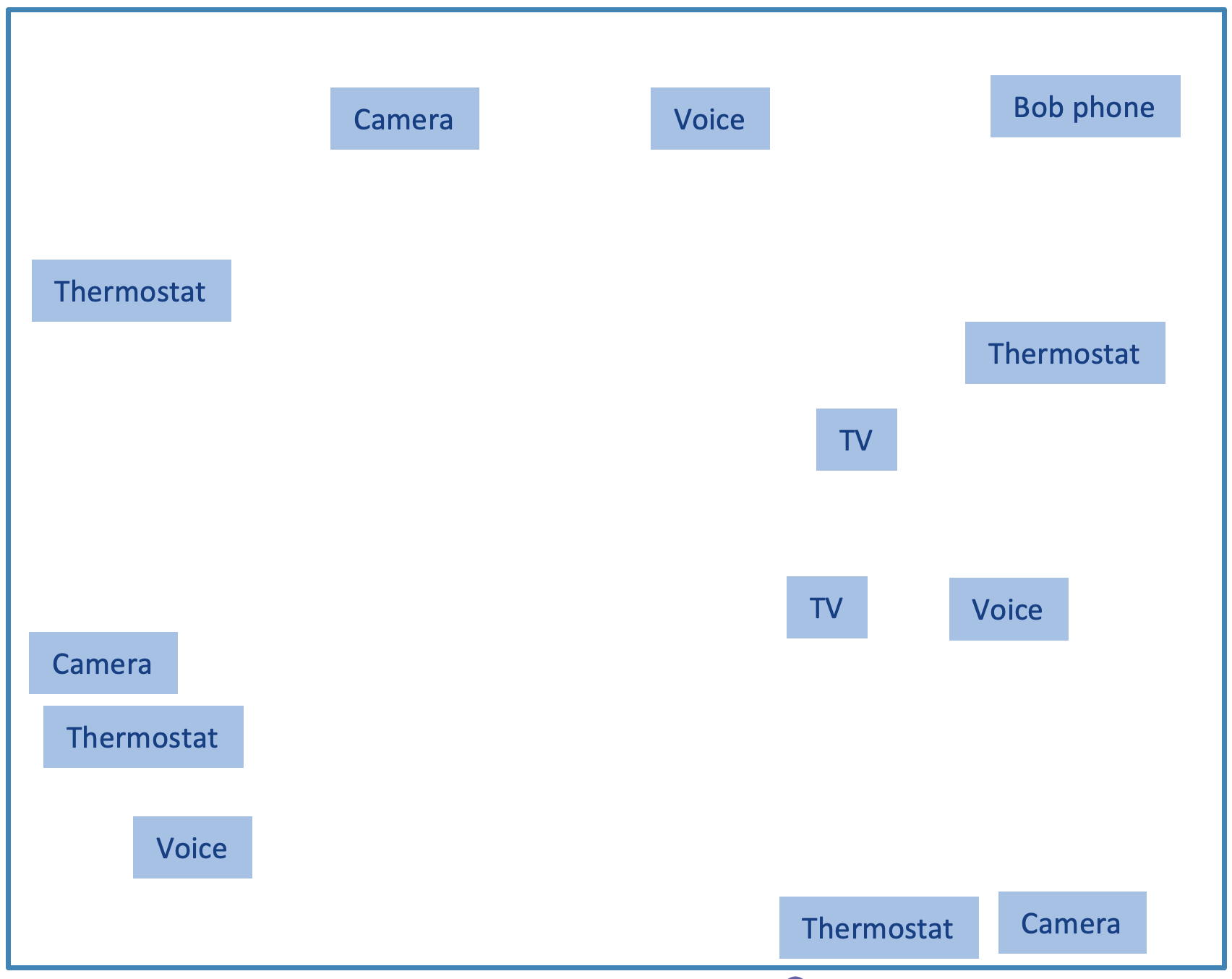}
         \caption{Map}
         \label{fig:devmap}
     \end{subfigure}
\caption{Managing home devices. (a) A typical app would present the home devices in a list view. (b) We propose presenting the home devices in a map view, showing where each device is located in the home.}
\label{fig:manage}
\end{figure}

A traditional way for a Wi-Fi management application to present the home devices is in a list view (figure~\ref{fig:devlist}). Such a view can be inconvenient for the user: it is hard to identify which device is which (for example if your home has three voice control devices). We propose to present the home devices in a map view, showing \emph{where} each device is located in the home (figure~\ref{fig:devmap}).

\begin{figure}
     \centering
     \begin{subfigure}[t]{0.95\linewidth}
         \centering
         \includegraphics[width=\textwidth]{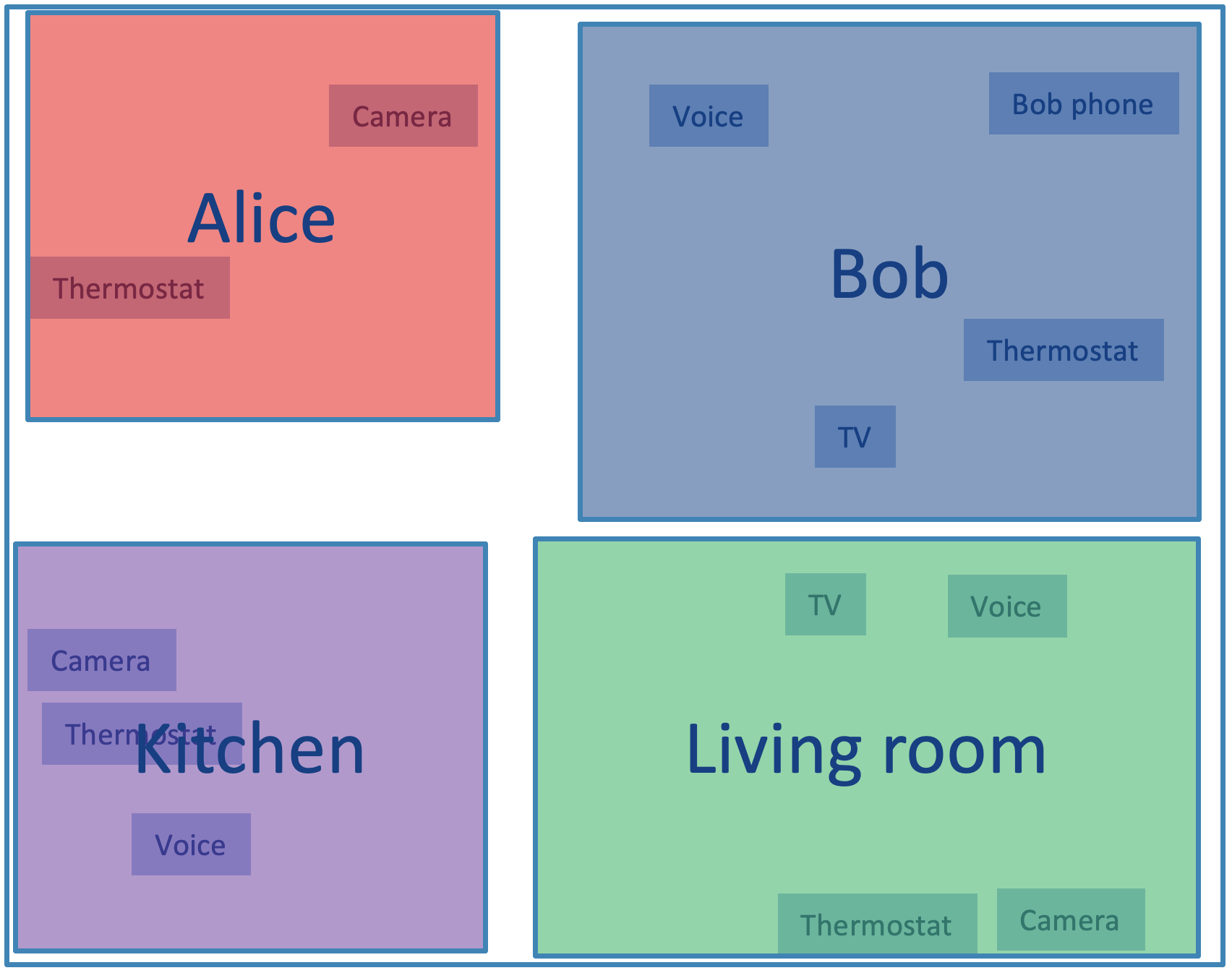}
         \caption{User profiles}
         \label{fig:profilemap}
     \end{subfigure}

    \begin{subfigure}[t]{0.95\linewidth}
         \centering
         \includegraphics[width=\textwidth]{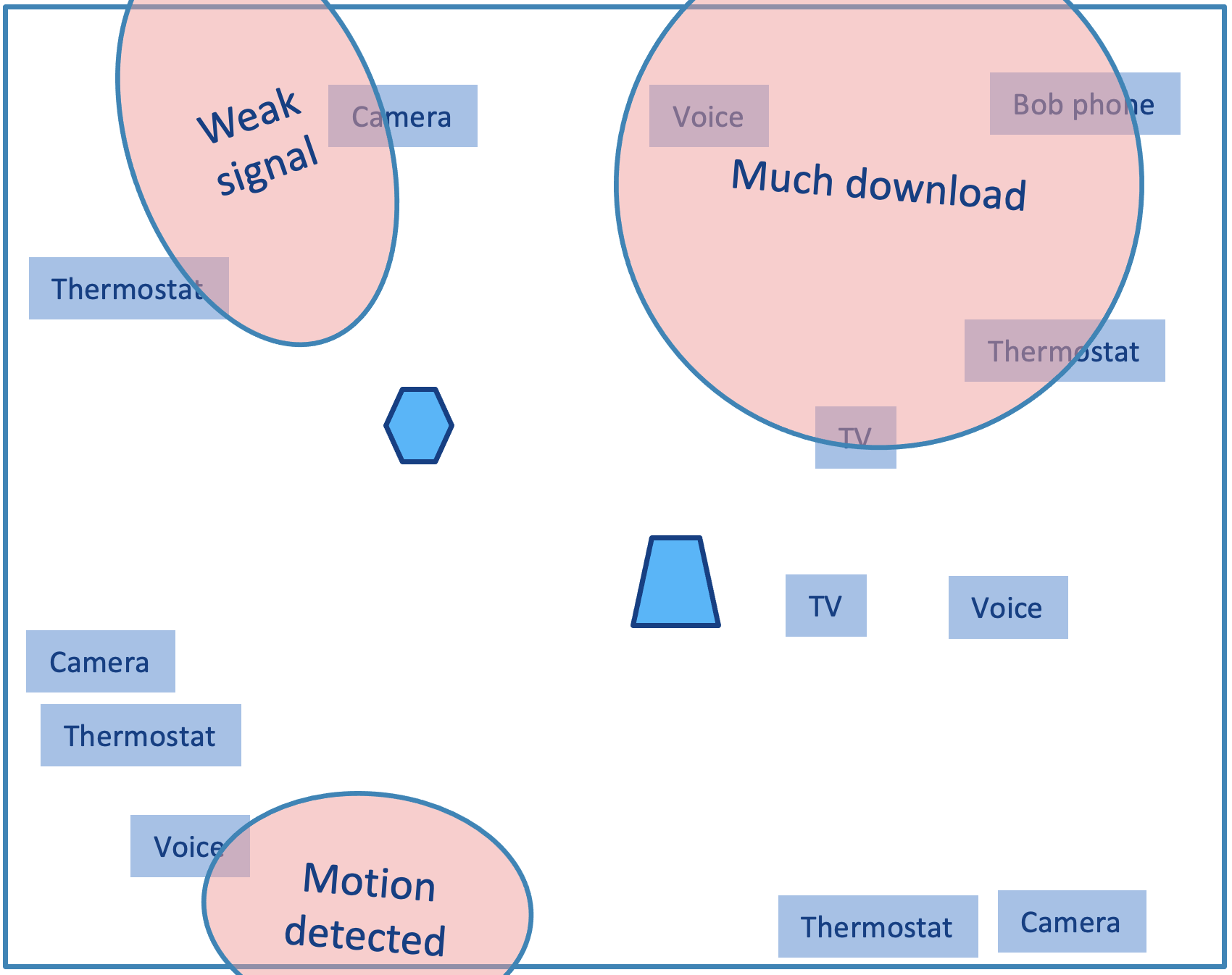}
         \caption{Wi-Fi tracking}
         \label{fig:Wi-Fimap}
     \end{subfigure}
\caption{Applications of a map view. (a) With a map view it is easier to identify devices that belong to the same person, and group them into a user profile. (b) A map view can help track Wi-Fi activity and quality in the home, for example, areas in the home with much download activity, or areas that consistently suffer from weak signal. This can help figure out the best place to put the access point (trapezoid) or places to put Wi-Fi extenders (hexagon). Methods that use Wi-Fi signals to detect motion can benefit from the map and better describe where the motion is detected.}
\label{fig:manage_applications}
\end{figure}

A map view of the devices in the home will provide a convenient user interface to manage and control devices in the home. Before we explain \emph{how} to create the map (the MapiFi method), we cover several features for customer experience improvement and troubleshooting, showing \emph{why} such a map is useful:

\textbf{Device identification and user profiling.}
A home may have multiple web-cameras, multiple voice-control devices, multiple smart TVs,~\etc\ When looking at a list view of the devices (figure~\ref{fig:devlist}), it can be hard to identify which device is which. A map view (figure~\ref{fig:devmap}) can resolve this ambiguity --- the user can clearly distinguish between the TV in the living room and the TV in the bedroom (see illustration in figure~\ref{fig:profilemap}). Similarly, viewing all the devices on a map can help group devices into user-profiles based on functional space (\eg~grouping the camera and thermostat located in Alice's room into a user-profile for Alice (Figure~\ref{fig:profilemap})). Such user profiles can then be useful for tracking internet usage by user, or for control features like parental control (\eg~easily pausing internet access at night for all the child's devices).

\textbf{Monitoring Wi-Fi activity and quality.}
Users are often interested in tracking how much internet they are using (\eg~to make sure they have the appropriate internet service plan, to track how much time they spend online gaming,~\etc). With a map view, the app can present internet activity (\eg~how much download) as a heat map --- this can help the user see which areas of the house (possibly associated with a certain person or family activity) use much traffic (figure~\ref{fig:Wi-Fimap}). It can also help detect unusual internet activity.
The user can also visualize Wi-Fi quality measures (like signal strength or the rate of corrupted packets) as a heat map on the map of the home --- this can help detect areas in the house that consistently get poor Wi-Fi conditions.

\textbf{Recommending gateway and Wi-Fi extenders placement.}
Based on the Wi-Fi signal coverage around the house, mapping the devices can help decide on the best place to put the main access point (the home's internet gateway, illustrated by the trapezoid in figure~\ref{fig:Wi-Fimap}). This can either be done by the user looking at the map view with the visualized signal strength and deciding on their own, or by some automated algorithm that takes into account devices locations, their signal strengths and amount of traffic to optimize gateway placement. Similarly, such algorithms can recommend the user to add Wi-Fi extenders (illustrated by the hexagon in figure~\ref{fig:Wi-Fimap}), and where to put them, typically in locations leading to a place in the home that consistently gets weak signal.

This analysis can be done on the day of installing a new Wi-Fi network, or after a few weeks or months of the residents using the Wi-Fi network. Visualizing signal strength on the map can be a helpful tool to explain to the user why they experience Wi-Fi trouble in some areas of the house or to convince them to change the gateway position or install Wi-Fi extenders.

\textbf{Localizing motion.}
There's a growing research area about using Wi-Fi radio signals to detect motion in the home~\cite{youssef2007challenges,gong2015wifi}. Methods usually rely on tracking the RSSI (received signal strength indicator) of signals coming from devices, or more detailed measurements like CSI (channel state information). When there is no motion in the home, typically the RSSI or CSI from devices should be stable. On the other hand, when someone is moving in a room, their body affects the obstruction and reflection of the radio signals in the room, causing fluctuations in the measurements. When we know the location of all the home devices, an algorithm can analyze the signal changes measured from all the devices and infer the location of the motion in the home. For example (figure~\ref{fig:Wi-Fimap}) if the gateway (the trapezoid in the center of the house) senses co-occurring fluctuations in signals from the three devices in the kitchen (lower left quadrant of the map), but stable signals from the other devices in the house, the algorithm can infer that there is motion in the kitchen.

\textbf{Locating a lost device.}
The features mentioned so far are mainly practical for a map of the stationary devices in the home --- devices that typically stay in their fixed location (TV, voice control, thermostat, printer,~\etc). However, MapiFi can also help with mobile devices, like tablets --- it can help locate a device that you lost in your home (if the device is still connected to the home network). Whenever you lose a device, you can use the MapiFi method (including a new ``walk around'' to take measurements, and then running the localization algorithm) to produce an ad-hoc map of your home. Then you'd see all the devices in the home --- some of them will be the stationary devices (which you may have already mapped before), and one of them will be the lost device --- the map will tell you where the device is.

In the next section, we describe the MapiFi method.

\section{Method}
\label{sec:methods}
The MapiFi method has two parts: taking measurements~\ref{ssec:taking_measurements} (which should take a few minutes) and a localization algorithm~\ref{ssec:localization_algorithm} (taking a few seconds). A typical time to use MapiFi is after connecting all the home devices to a new Wi-Fi network --- this will produce a map with the locations of all the stationary devices in the home (the user can decide to ignore/delete locations of mobile devices, like phones, if they assume that these devices will keep moving around the house). The same map can be used for a long time, possibly months or years. The user can use MapiFi again whenever they want, if they added new devices to the home, or if they re-arranged the locations of devices in the home.
Additionally, if some day someone loses a device in the house (\eg~someone forgot where they left their tablet), they can use MapiFi again, to produce the current map and see where their lost device is.

\subsection{Taking measurements}
\label{ssec:taking_measurements}

The first part of MapiFi requires participation from the user. This can be the end user of the Wi-Fi network (the resident of the home) or a technician coming to help troubleshoot the home network. The user carries a mobile device (``the measuring device'') and walks around with it in the home, pausing for a few seconds in multiple points along the path --- we refer to these as ``anchor'' points. 
In each anchor point, the user runs a script that listens to the home network's Wi-Fi channel and captures Wi-Fi packets coming in from all the home devices. The measuring device ignores the content of the packets, and only registers the MAC (media access control) address of the sending device and the RSSI of each packet.

It doesn’t matter which channel the home network is using. It is fine if the network changes channels during the walk (as long as the measuring device knows which channel to listen to). The key thing is to catch incoming packets from the devices and measure the signal strength (as a proxy for distance). It is also possible to ``ping'' the home devices (\eg~send some HTTP request to the smart TV, either from the measuring device or from another collaborating device) in order to ``wake them up'' and make them reply and send something over Wi-Fi --- the point is to induce the devices to transmit anything over the air, so that the measuring device can measure signal strength.

The user can make the walking path longer, more winding, exploring all rooms, with more anchor points, and can hold the measuring device in different heights. All these factors can help gather more diverse evidence (measurements), which will help the localization algorithm estimate device locations more accurately.

\subsection{Localization algorithm}
\label{ssec:localization_algorithm}

\subsubsection{From signal-strength to range}
\label{sssec:range}
MapiFi relies on the fact that radio signals lose power as they travel through the air: generally, we will sense signals coming from near devices with higher power than signals coming from far devices. In practice, the level of signal attenuation depends on the frequency band, obstacles along the way, reflection paths,~\etc\ We start simple by assuming a constant decay factor, meaning that the power of the signal reduces (from the transmit power $Tx$ to the measured, received power $Rx$) by a constant power $\gamma$ of the distance (or ``range'' $r$) that it travelled (similar to previous models of signal decay through space,~\eg~\cite{shang2014location}). 
\begin{align}
r^\gamma &= \frac{Tx}{Rx}
\end{align}
or, if the powers are in decibel:
\begin{align}
r^\gamma &= 10^{\frac{TxdB - RxdB}{10}}
\end{align}
An important part of MapiFi is the realization that every device may transmit signals in a different transmit power. To simplify the method, we calculate the range of a signal assuming a fixed arbitrary transmit power ($TxdB=-50dB$) and we model these inter-device differences later --- by a distance-gain variable.
We assumed a constant decay factor of $\gamma=2.5$. So, for every measurement, we use the RSSI $RxdB$ to calculate the range of the signal:
\begin{align}
r &= 10^{\frac{TxdB - RxdB}{10\gamma}} \\
  &= 10^{\frac{-50 - RxdB}{10*2.5}}
\end{align}
We then treat the range $r$ as an uncalibrated measure of the distance between the measuring (anchor) point and the transmitting device. We model the differences among devices with a multiplicative gain, $g$, and we assume that every device $j$ has a fixed gain $g_j$: so the distance between device $j$ and anchor $i$ is modeled as $g_j r_{i,j}$.

This simplified model assumes that the device stays in its place during the few minutes of walk-around and that it is trying to communicate with the access point, which also stays in its place. Remember, even if the measuring device keeps moving, it doesn’t matter, because the packets are not targeted to it, they are always targeted to the access point. It is possible that during the walk there is suddenly more noise, and the home device increases its transmitting power. Also, if during the walk the home device switches Wi-Fi frequency band from 2.4GHz to 5GHz then its signals will attenuate more quickly with distance. In such cases, the fixed-distance-gain assumption may fail. Future methods can mitigate these issues by adding variables to the model, or by helping the measurement walk be quicker and more efficient. 

This model describes the distance that the signal travels from the transmitting device to the measuring device, which can be a non-straight path. In this preliminary MapiFi version, we further simplify assumptions and treat it as an approximation of the direct distance in 3d space between the transmitting and measuring devices. Next, we describe how the localization algorithm uses all these distances to estimate the location of every device (and every anchor point) in the 3d space of the home.

\subsubsection{Finding a device's location and gain}
\label{sssec:device_location}
In the basic problem, we assume we know the $N$ anchor locations $a_i\in \mathbb{R}^3$ for $i \in\{0,\ldots,N-1\}$, and we have the signal strength measurements (which we convert to ranges $r_i\in \mathbb{R_+}$ between the device and all the $N$ anchors). We need to find out the device's location $d\in \mathbb{R}^3$ and its gain $g\in \mathbb{R_+}$. The evidence (measurements) gives us a set of $N$ quadratic equations in $d$ and $g$:

\begin{align}
& \forall i \in\{0,\ldots,N-1\}: \\
& || d - a_i ||_2^2 = (gr_i)^2
\end{align}
or
\begin{align}
& \forall i \in\{0,\ldots,N-1\}: \\
& ||d||_2^2 + ||a_i||_2^2 - 2a_i^Td = r_i^2g^2
\end{align}

By subtracting equations $i\in\{1,\ldots,N-1\}$ from equation $0$, we get a system of $N-1$ linear equations (in $d$ and $g^2$):
\begin{align}
& \forall i \in\{1,\ldots,N-1\}: \\
& ||a_0||_2^2 - ||a_i||_2^2 = 2(a_0-a_i)^Td + (r_0^2-r_i^2)g^2
\end{align}
With enough equations (enough anchor points, in general position), this system can theoretically be solved for the device location $d$ and its gain $g$ (total of 4 unknown variables) --- it would require 4 linearly independent equations, which would require 5 quadratic equations. This means that in perfect conditions, 5 anchor points will suffice to fully recover the device's location and gain. However, because the measurements are noisy, these equalities can be approximated by solving a least squares problem, possibly with constraints ($g$ needs to be positive, and the coordinates of $d$ can be constrained by the dimensions of the house, if these are readily available). Having more than 5 anchor points would provide more evidence (the more, the better) and help better infer the device's location. If the anchor points (the locations where the user stops to take measurements) are on a straight line, it may result in redundant (linearly dependent) equations --- this is why it is best to take a winding path through the home, as well as to hold the measuring device at different heights.

\subsubsection{Finding an anchor's location}
\label{sssec:anchor_location}
In the complementary problem, we assume that we know the locations $d_j\in \mathbb{R}^3$ and gains $g_j\in \mathbb{R_+}$ of all the $M$ home devices ($j\in\{0,\ldots,M-1\}$). From the measurements, we have the ranges $r_j\in \mathbb{R_+}$ between all the $M$ devices and a specific anchor, and we need to find out this anchor's location $a$. The evidence gives us a set of $M$ quadratic equations in $a$:

\begin{align}
& \forall j \in\{0,\ldots,M-1\}: \\
& ||a - d_j||_2^2 = r_j^2g_j^2
\end{align}
or
\begin{align}
& \forall j \in\{0,\ldots,M-1\}: \\
& ||a||_2^2 + ||d_j||_2^2 - 2a^Td_j = r_j^2g_j^2
\end{align}
As in the previous problem, we can turn this to a system of $M-1$ linear equations in $a$, and treat this as a least squares problem, with constraints on $a$'s coordinates to be inside the house (if the house's coordinate boundaries are available).

\subsubsection{Full algorithm}
\label{sssec:mapifi_algorithm}
In some cases the locations of the anchors (the measurement points) can be known in advance --- either if the measurements are taken from multiple stationary network devices (like Wi-Fi extenders), or if the user carefully recorded the coordinates of every point along the path where they took measurement. In such rare cases, you can simply solve the first problem for each home-device separately to get its location.

However, in the general case, the anchor locations are not known because the user walks freely in a winding path through the house. So in the general case, we have $N$ unknown anchor locations $a_i$ for $i\in\{0,\ldots,N-1\}$, $M$ unknown device locations $d_j$ and gains $g_j$ for $j\in\{0,\ldots,M-1\}$. The observed variables are the measured ranges $r_{i,j}, i\in\{0,\ldots,N\}, j\in\{0,\ldots,M\}$ (where for some of the anchor-device pairs we are missing a measurement).

The solution is an alternating algorithm: start with a random initialization (guess) of the anchor locations $a_i$, and then iterating over the steps:
\begin{enumerate}
\item estimating the device locations $d_j$ and gains $g_j$ given the currently assumed anchor locations $a_i$ (\ref{sssec:device_location}),
\item estimating the anchor locations $a_i$ given the currently assumed device locations $d_j$ and gains $g_j$ (\ref{sssec:anchor_location}),
\item (optional) normalizing the estimated anchor locations. Make sure the points don't explode in space or decay to the origin (or make sure they remain within a desired boundary box).
\end{enumerate}

\section{Initial experiment}
\begin{figure}
\begin{subfigure}[t]{\linewidth}
  \includegraphics[width=\linewidth]{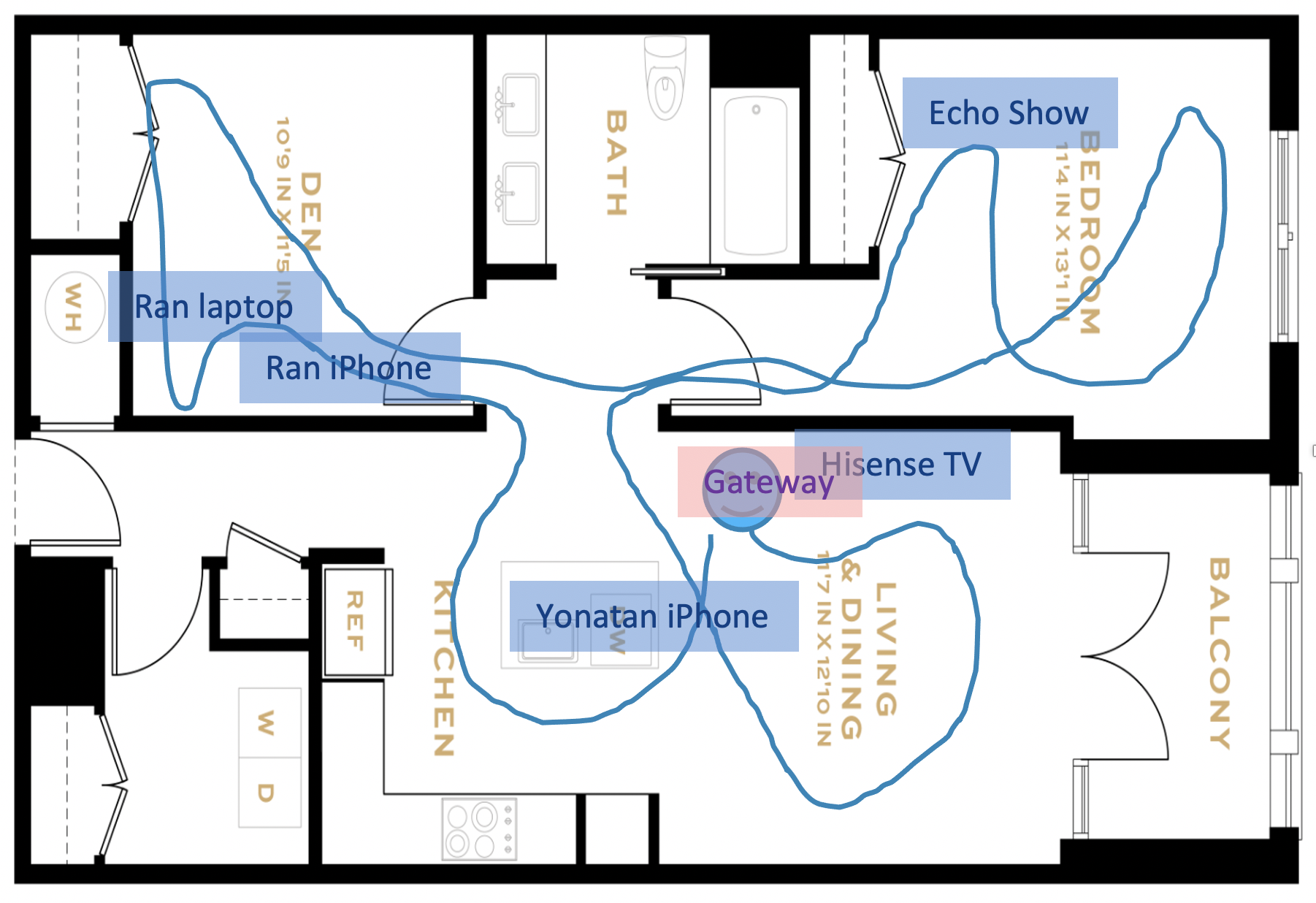}
  \caption{Taking measurements in a real home.}
  \label{fig:floor_plan}
\end{subfigure}

\begin{subfigure}[t]{\linewidth}
  \includegraphics[width=\linewidth]{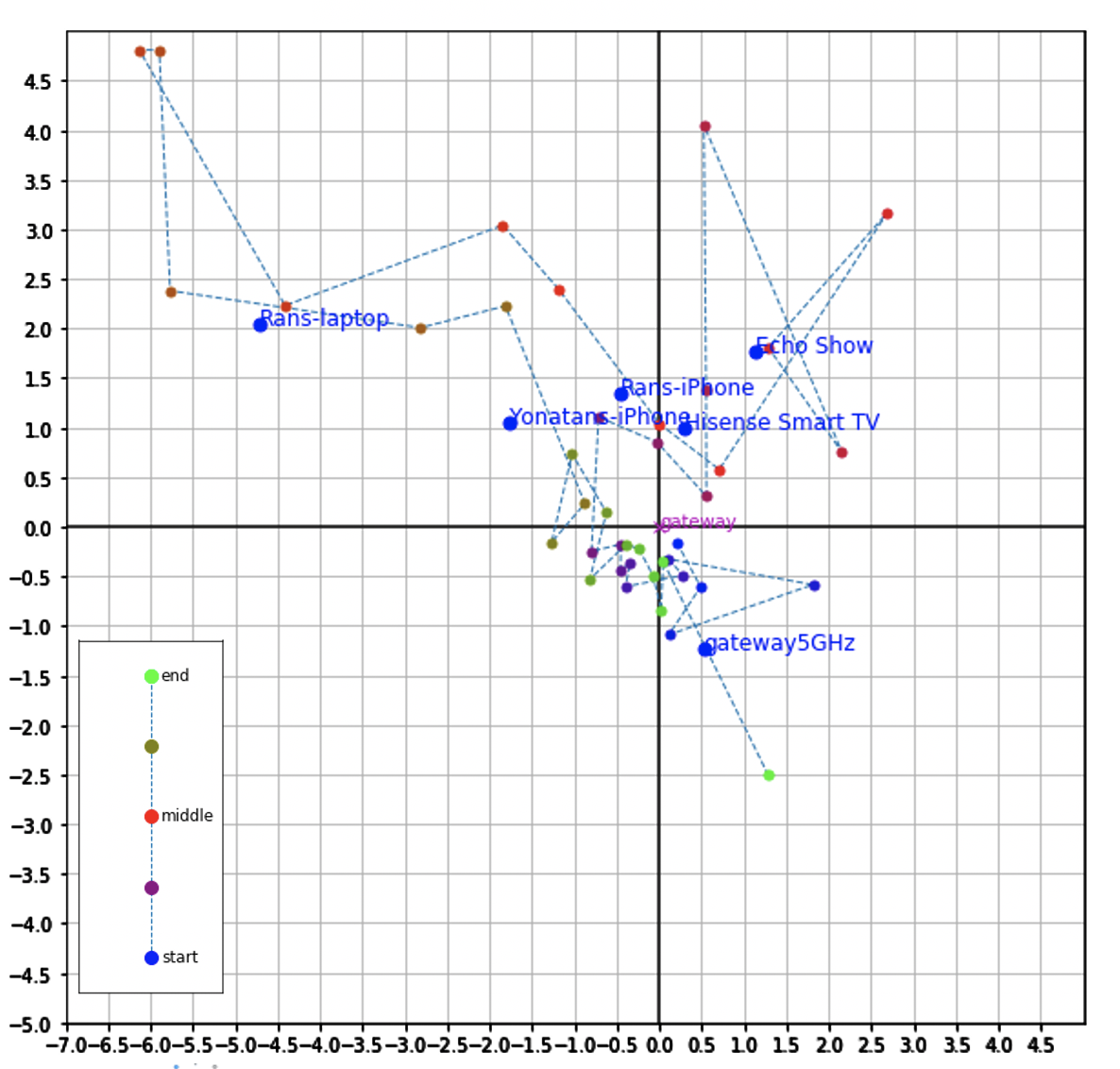}
  \caption{MapiFi: the resulting map. Projection to 2d (ignoring the height coordinate). The estimated measurement path is marked with dashed line with anchors as small dots with colors ranging from blue at the start of the path to green at the end of the path. The devices are marked with bold blue dots at their estimated locations.}
  \label{fig:finalmap}
\end{subfigure}

\caption{Initial experiment. a) The walk through a real home to take measurements. b) The resulting map.}
\label{fig:experiment}
\end{figure}

We tested the MapiFi method in one initial experiment, in an actual single-floor apartment with one bedroom and one den (see figure~\ref{fig:floor_plan}). The internet gateway and five devices were placed in known locations. We used a laptop as the measuring device. The user walked with the laptop in a path through the rooms (marked in blue curve in the figure), stopped in 40 points along the path (the ``anchor'' points) and took measurements (signal strengths of Wi-Fi packets from the devices in the home). This walk took approximately five minutes.

The localization algorithm managed to reconstruct an estimated path that resembles the actual path that the user took: starting close to the gateway, going through the bedroom (close to the ``Echo''), passing through the den (close to ``Ran's laptop'' and ``Ran's iPhone''), then passing by ``Yonatan's iPhone'' (see figure~\ref{fig:finalmap}). The estimated locations of the devices approximate the relative locations of the actual devices in the home.

\begin{figure}
     \centering
     \begin{subfigure}[t]{0.3\linewidth}
         \centering
         \includegraphics[width=\textwidth]{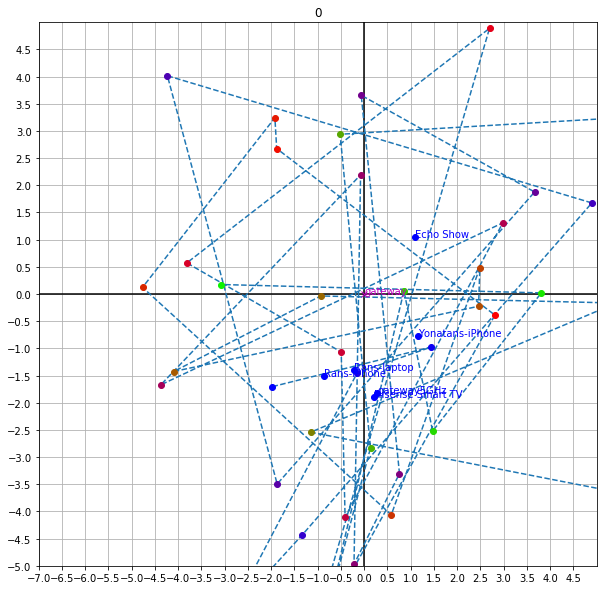}
         \caption{Iteration 0}
         \label{fig:iter0}
     \end{subfigure}
    \begin{subfigure}[t]{0.3\linewidth}
         \centering
         \includegraphics[width=\textwidth]{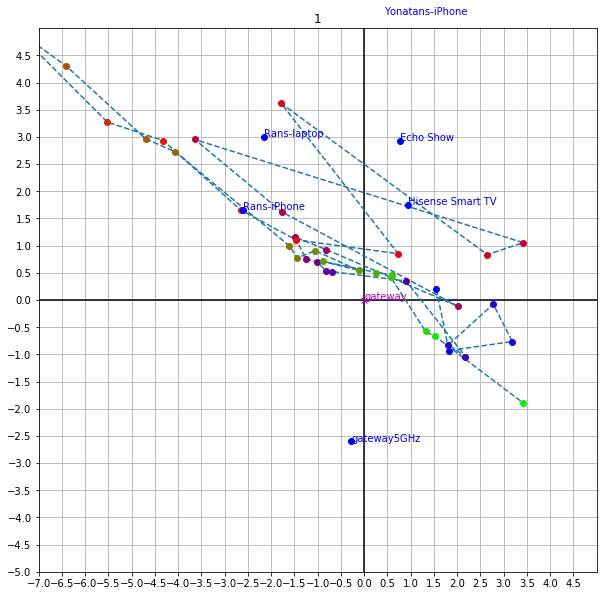}
         \caption{Iteration 1}
         \label{fig:iter1}
     \end{subfigure}
    \begin{subfigure}[t]{0.3\linewidth}
         \centering
         \includegraphics[width=\textwidth]{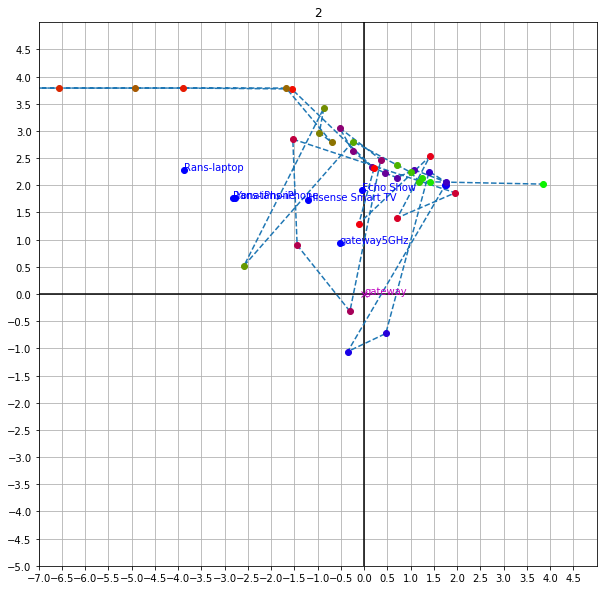}
         \caption{Iteration 2}
         \label{fig:iter2}
     \end{subfigure}

    \begin{subfigure}[t]{0.3\linewidth}
         \centering
         \includegraphics[width=\textwidth]{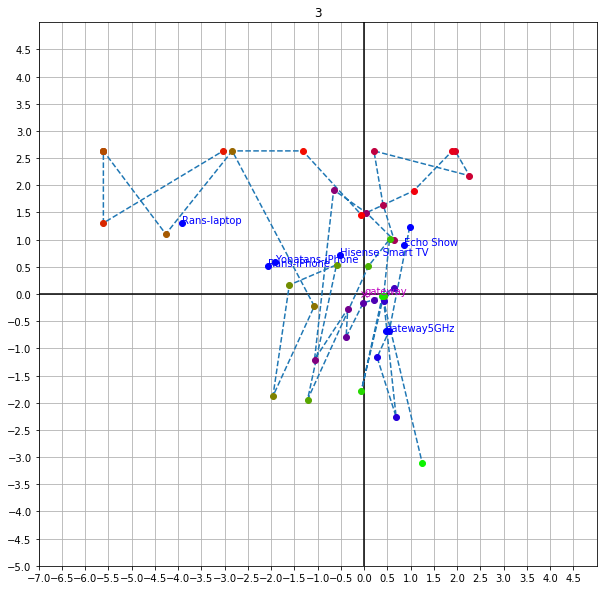}
         \caption{Iteration 3}
         \label{fig:iter3}
     \end{subfigure}
    \begin{subfigure}[t]{0.3\linewidth}
         \centering
         \includegraphics[width=\textwidth]{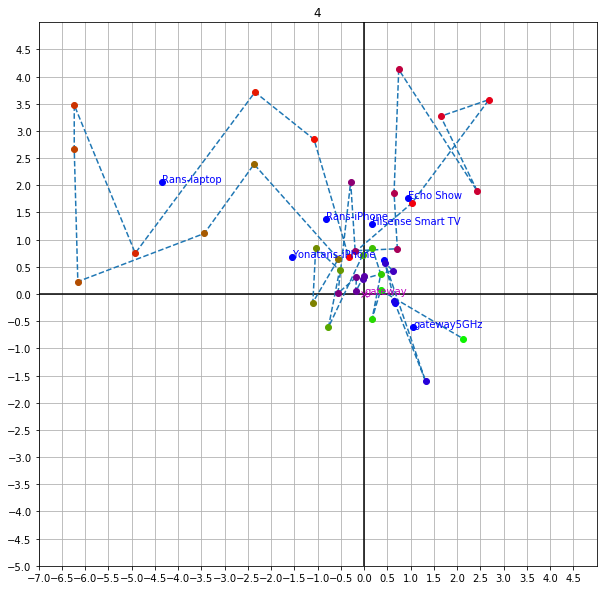}
         \caption{Iteration 4}
         \label{fig:iter4}
     \end{subfigure}
    \begin{subfigure}[t]{0.3\linewidth}
         \centering
         \includegraphics[width=\textwidth]{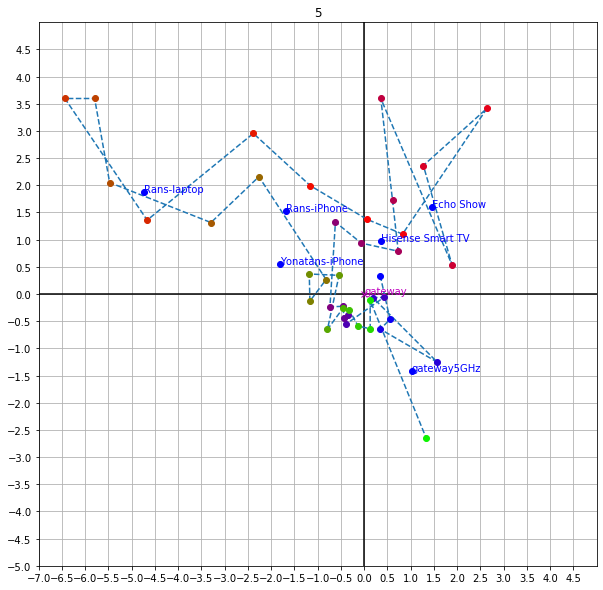}
         \caption{Iteration 5}
         \label{fig:iter5}
     \end{subfigure}

\caption{Iterations of MapiFi localization. The algorithms starts with random initialization of the anchors along the path. As the iterations progress, the estimated path becomes more coherent and resembles the actual path that the user took in the home. The devices' estimated locations change with every iteration, following the estimated locations of the anchors.}
\label{fig:iterations}
\end{figure}

Figure~\ref{fig:iterations} shows what happens in the algorithm: starting from random initialization, and evolving to better estimates of both the measurement path (the anchor points) and the devices locations.

\section{Conclusions}
\label{sec:conclusions}
We describe a method for mapping Wi-Fi devices in a home's space. The generated map can help the resident manage their home devices and monitor their Wi-Fi quality and usage. We report promising results from an initial experiment in a real home environment. The method itself can be further improved with various adjustments, like combining known device-locations as constraints, giving stronger weight to measurements of closer-devices (where distance estimation is more reliable), modeling obstacles (walls, furniture,~\etc), and combining with a structural model of rooms and walls from computer-vision based methods.

%


\vfill\pagebreak


\bibliographystyle{IEEEbib}
\bibliography{refs}

\end{document}